\documentclass[twocolumn,showpacs,prl,superscriptaddress]{revtex4}

\usepackage{graphicx}
\usepackage{bm}
\usepackage{amsmath}
\usepackage{amssymb}

\begin{document}

\title{Giant anisotropy of Zeeman splitting of quantum confined
acceptors in Si/Ge}

\author{K.-M.~Haendel}
\affiliation{Institut f\"ur Festk\"orperphysik, Universit\"at
Hannover, Appelstrasse 2, 30167 Hannover, Germany}
\author{R.~Winkler}
\altaffiliation[Present adress: ]{Department of Physics, Northern
Illinois University, De Kalb, IL 60115}
\affiliation{Institut f\"ur Festk\"orperphysik, Universit\"at
Hannover, Appelstrasse 2, 30167 Hannover, Germany}
\author{U.~Denker}
\author{O.~G.~Schmidt}
\affiliation{Max-Planck-Institut f\"ur Festk\"orperforschung,
Heisenbergstrasse 1, 70569 Stuttgart, Germany}
\author{R.~J.~Haug}
\affiliation{Institut f\"ur Festk\"orperphysik, Universit\"at
Hannover, Appelstrasse 2, 30167 Hannover, Germany}

\date{\today}

\begin{abstract}
  Shallow acceptor levels in Si/Ge/Si quantum well heterostructures
  are characterized by resonant tunneling spectroscopy in the
  presence of high magnetic fields. In a perpendicular
  magnetic field we observe a linear Zeeman splitting of the
  acceptor levels. In an in-plane field, on the other hand, the
  Zeeman splitting is strongly suppressed. This anisotropic Zeeman
  splitting is shown to be a consequence of the huge light
  hole-heavy hole splitting caused by a large biaxial strain and a
  strong quantum confinement in the Ge quantum well.
\end{abstract}



\maketitle

Spintronic and quantum computing \cite{aws02,los98} are novel device
concepts relying on quantum mechanical coherence. Si/Ge-based
systems are promising candidates offering long spin coherence times
\cite{tyr05, tyr03}, fast operations, and a well-established record
of scalable integration. These important properties are also crucial
requirements \cite{los98, pre98, sou03} for implementing multi-qubit
operations in a future quantum computer. One concept that may form
the technological basis of a quantum computer is the spin-resonance
transistor (SRT) \cite{kan98}. Vrijen \textit{et.\ al.}\
\cite{vri00} proposed an SRT where the electron spin manipulation is
realized using the change in $g$-factor between Si-rich and Ge-rich
environments of a Si/Ge-heterostructure. However, engineering the
$g$-factor in such systems is complicated by the fact that the
electron states in Si are in the $X$ valleys whereas in Ge the
electrons are located in the $L$ valleys \cite{bar03}. This problem
does not arise for the valence band states, as both Si and Ge have
their valence band maximum at the $\Gamma$ point. Thus, valence band
states in Si/Ge are a promising choice for $g$-factor engineering in
search for spin manipulation.

In this paper we have analyzed the $g$-factor of shallow acceptor
levels in a Si/Ge-heterostructure by resonant-tunneling
spectroscopy. We find that their effective $g$-factor is highly
anisotropic, giving a large Zeeman splitting of the acceptor states
in a perpendicular field, whereas we cannot resolve any Zeeman
splitting in in-plane fields up to 18~T. This giant anisotropy
provides the possibility to tune the coupling of the holes to an
external magnetic field by a gate-controlled shift of the hole wave
function \cite{mal01} in spintronic devices.

\begin{figure}[h!]
  \includegraphics[width=.45\textwidth]{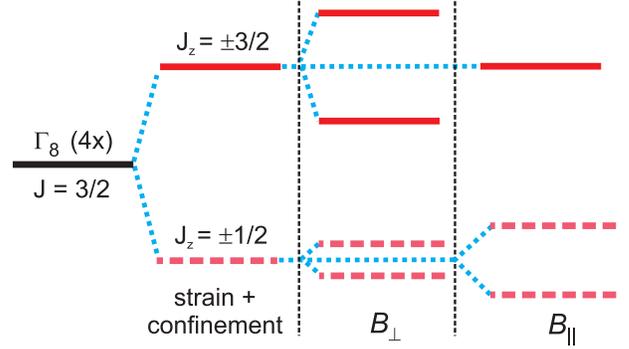}
  \caption{Qualitative sketch of the valence band states at different
  perturbation conditions. Already at $B=0$, biaxial strain and a
  confinement potential reduce the fourfold degeneracy to twofold,
  giving heavy-hole (HH) states with $J_{z}=\pm 3/2$ and light-hole
  (LH) states with $J_{z}=\pm1/2$. Depending on the orientation of
  the magnetic field, the twofold degeneracies of the HH and LH
  states will be lifted or preserved.}
  \label{fig1}
\end{figure}

For a proper understanding of acceptor levels it is essential to
take into account the fourfold degeneracy of the valence band at the
$\Gamma$ point (Fig.~\ref{fig1}) which reflects the fact that the
bulk valence band edge in these materials is characterized by an
effective angular momentum $J = 3/2$ \cite{lut56, win03}. As the
symmetry of the crystal is reduced due to biaxial strain and a
confinement potential, the degenerate states split into heavy-hole
(HH) subbands with $J_{z} = \pm 3/2$ and light-hole (LH) subbands
with $J_{z} = \pm 1/2$. Here, the quantization axis for the angular
momentum is the $z$-axis perpendicular to the epitaxial layer. So
both parameters, the confinement potential and the built-in strain,
substantially influence the energy levels of an acceptor in a
quantum well (QW) \cite{mas85, ales02}. In a magnetic field
$B_{\perp}$ orientated perpendicular to the epitaxial layer we get a
Zeeman splitting of HH and LH states, $\Delta E_\mathrm{HH \,
(LH)}^\perp = g_\mathrm{HH \, (LH)}^{\perp} \mu_\mathrm{B}
B_{\perp}$, where $g_\mathrm{HH \, (LH)}^\perp$ is the $g$-factor of
the HH (LH) states in a perpendicular field and $\mu_\mathrm{B}$ is
the Bohr magneton. But for an in-plane magnetic field $B_{\|}$ the
linear Zeeman splitting of HH states is suppressed because there is
no $B_\|$-induced direct coupling between these states, $\langle
\mathrm{HH}| \bm{J} \cdot \bm{B}_{\|} | \mathrm{HH} \rangle = 0$,
where $\bm{J}$ is the vector of $J=3/2$ spin matrices
\cite{kes90,gla90,lin91,gol93}. This does not apply for LH states
which show a significant Zeman splitting $\Delta E_\mathrm{LH}^{\|}
> 0$. We emphasize that the vanishing Zeeman splitting of HH states
in a parallel field reflects the fact that the HH-LH splitting in
our samples is much larger than the maximal Zeeman energies ($\sim
7$~meV) \cite{win03}.

\begin{figure}[t]
  \includegraphics[width=.45\textwidth]{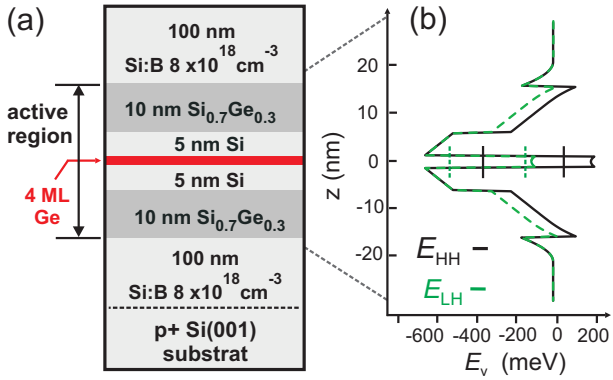}
  \caption{(a) Layer sequence of
  the heterostructure. (b) Self-consistently calculated valence band
  profile of the active region. The solid line shows the shape of
  the heavy-hole (HH) subband and the dashed line the light-hole
  (LH) subband. Due to biaxial strain and a strong quantum
  confinement a huge HH-LH splitting results.}
 \label{fig2}
\end{figure}

Figure~\ref{fig2} shows the layer sequence and the valence band
profile of our samples. They are prepared by growing a 100~nm thick
Si:B layer ($p = 8 \times 10^{18}$~cm$^{-3}$) on a (001) $p^{+}$-Si
substrate. The active region of the samples consists of two 10~nm
thick SiGe QWs separated by a 10~nm thick Si barrier. In the center
of the Si barrier a 4~monolayers (ML) thick Ge QW is embedded.
Finally, the active region is capped with 100~nm Si:B ($p = 8 \times
10^{18}$~cm$^{-3}$). For the dc-transport measurements we have
fabricated diodes with lateral diameters of 1~$\mu$m. Measurements
were performed at temperatures down to $T= 50$~mK and using magnetic
fields up to 18~T.

\begin{figure}[t]
  \includegraphics[width=.4\textwidth]{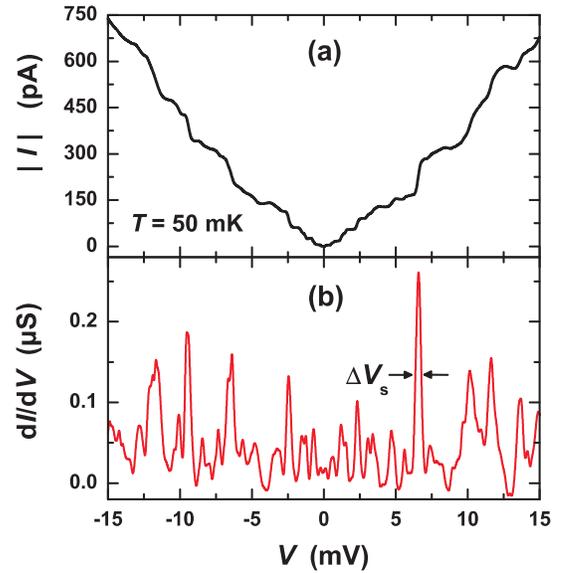}
  \caption{(a) Current--voltage ($I-V$) characteristics of a
  diode with a diameter of 1~$\mu$m at $T= 50$~mK and (b) the
  corresponding differential conductance ($dI/dV$). A current step
  respectively a differential conductance peak occurs whenever
  resonant tunneling through a shallow acceptor level in the Ge
  quantum well is energetically possible.}
 \label{fig3}
\end{figure}

Figure~\ref{fig3}(a) shows a typical current-voltage ($I-V$)
characteristic of a diode at $T= 50$~mK. A staircase structure is
observed that is even better resolved in the differential
conductance ($dI/dV$) curve shown in Fig.~\ref{fig3}(b). The
steplike increase of the current is in a bias range which is about
300~mV below the onset of resonant tunneling of holes through the 2D
subbands of the central Ge QW so that this mechanism cannot explain
the current steps. We attribute these current steps in the pA-range
to tunneling processes of holes through zero-dimensional acceptor
levels of Boron dopant-atoms which have migrated into the Ge QW from
the highly doped Si:B contact regions.

A resonant tunneling process through an acceptor level $E_{s}$
occurs each time $E_s$ is in resonance with the Fermi energy
$E_\mathrm{F}$ of the emitter. The bias position of a current step
is given by $V_{s} = (E_{s} - E_\mathrm{F}) / \alpha e$, where
$\alpha$ is the bias-to-energy conversion coefficient. We determine
$\alpha$ from the temperature-dependent broadening of the current
step edges. As a measure of this broadening, we use the full width
at half maximum of the corresponding differential conductance peaks,
$\Delta V_{s}$ [see Fig.~\ref{fig3}(b)]. It increases according to
$(\alpha e\Delta V_{s})^{2} = (\Delta E_{s})^{2} + (3.53 \,
kT)^{2}$, where the term $3.53 \, kT$ stems from the broadening of the
Fermi function characterizing the carrier distribution. Using this
equation we obtain $\alpha = 0.5\pm 0.1$ for several peaks in both
bias polarities.

\begin{figure}[t]
  \includegraphics[width=.35\textwidth]{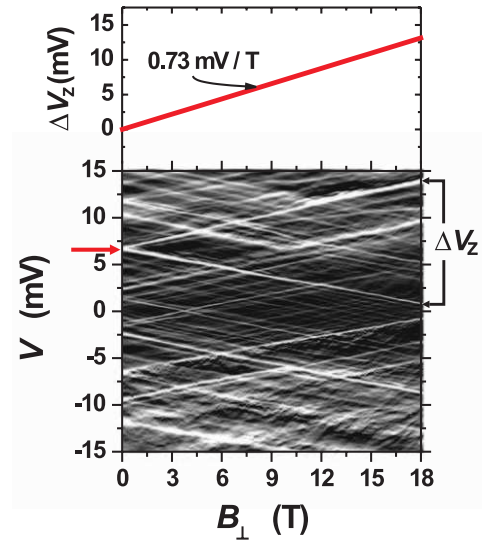}
  \caption{Gray-scale plot of the differential conductance at 50~mK
  for a magnetic field orientated perpendicular to the epitaxial
  layer, where a dark (bright) shade corresponds to small (large)
  conductance. Exemplarily, the top graph shows the splitting of a
  level, marked by an arrow at 6.5~mV.}
  \label{fig4}
\end{figure}

Figure~\ref{fig4} shows a gray-scale plot of the differential
conductance $dI/dV$ as a function of an external magnetic field
orientated perpendicular to the epitaxial layer, $B_{\perp}$. In the
voltage range from $-15$~mV to 15~mV the conductance maxima exhibit
a linear splitting as a function of $B_{\perp}$. All levels show
accurately the same splitting, as indicated by the parallel
evolution of the conductance maxima. As an example, the upper panel
of Fig.~\ref{fig4} shows the splitting of the conductance maximum at
6.5~mV. The gradient of the splitting is $\mathrm{d}\Delta
V_\mathrm{Z} / \mathrm{d}B_{\perp} = 0.73$~mV/T. We attribute the
linear splitting to the Zeeman splitting of the HH sublevels,
$\Delta E_\mathrm{Z} = g^{\perp}_\mathrm{HH} \mu_\mathrm{B}
B_{\perp}$. The $g$-factor $g_\mathrm{HH}^{\perp} $ can be
determined using
 \begin{equation}\label{eq-3}
    g_\mathrm{HH}^{\perp}
    = \frac{\alpha e}{\mu_\mathrm{B}}
      \frac{\mathrm{d} \Delta V_\mathrm{Z}}{\mathrm{d}B_{\perp}}.
\end{equation}
With $\alpha = 0.5$ and $\mathrm{d}\Delta V_\mathrm{Z} / \mathrm{d}
B_{\perp} = 0.73$~mV/T we obtain $g^{\perp}_\mathrm{HH} = 6.3$. This
value agrees well with optically measured $g$-factors \cite{soe73,
bro79} of group-III impurities such as B in Ge. This confirms the
assumption that the observed levels belong to Boron dopant-atoms
which have diffused from the heavily doped contact regions into the
Ge QW.

\begin{figure}[t]
  \includegraphics[width=.35\textwidth]{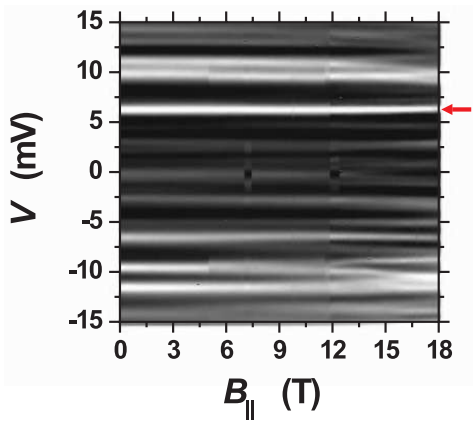}
  \caption{Gray-scale plot of the differential conductance at 1~K
  for a magnetic field orientated parallel to the epitaxial layer,
  where a dark (bright) shade corresponds to small (large)
  conductance.  The right arrow points to the same level as in
  Fig.~\ref{fig4}.}
  \label{fig5}
\end{figure}

Using the $B_\perp$ dependence of the conductance we can obtain an
upper limit for the radial extent $\lambda$ of the wave function of
a hole bound to an acceptor level. According to first-order
perturbation theory, a ground state acceptor level is affected by a
diamagnetic shift $\Delta E_\mathrm{D} \simeq e^{2}B_{\perp}^{2}
\lambda^{2} / 8m^\ast$, with $m^\ast \approx 0.28m_{0}$ the HH
effective mass of Ge. (We neglect here the modifications of the hole
wave functions caused by the confinement in the narrow QW.) But even
at $B_\perp = 18$~T a diamagnetic shift of the levels cannot be
observed in our measurements, only the contribution of the linear
Zeeman splitting can be seen in Fig.~\ref{fig4}. This implies that
the diamagnetic shift $\Delta E_\mathrm{D}$ of the acceptor level is
smaller than the width $\Delta E_{s} \simeq 0.15$~meV of the
conductance peak at 50~mK so that $\lambda$ must be smaller than
$2.5$~nm. Therefore, $\lambda$ of the acceptor wave functions in the
thin Ge QW is of the same order as the QW width. We remark that
Bastard \cite{bas81a} derived a simple model to estimate $\lambda$
for an impurity in a narrow QW which yields for our system
parameters $\lambda \approx 0.8 a_\mathrm{B}^\ast$, where
$a_\mathrm{B}^\ast \simeq 3.1$~nm is the effective Bohr radius for
heavy holes in Ge.

Next we present in Fig.~\ref{fig5} our results for the measured
conductance in an \emph{in-plane} magnetic field $B_\|$. While we
saw in Fig.~\ref{fig4} that $B_\perp$ gives rise to a significant
Zeeman splitting of the acceptor levels linear in $B_\perp$, it is
most remarkable that up to 18~T most conductance maxima are not at
all influenced by an in-plane magnetic field $B_{\|}$. The
conductance maximum at 6.5~mV, which exhibits a pronounced linear
splitting for $B_{\perp}$ (arrow in Fig.~\ref{fig4}), does not show
any splitting in the case of an in-plane magnetic field $B_{\|}$
(arrow in Fig.~\ref{fig5}). If a splitting exists, it must be
smaller than the width of the conductance peak which is about
0.35~meV \cite{temprat} here.

The giant anisotropy of the Zeeman splitting is a consequence of the
effective spin $J=3/2$ of the valence band states (Fig.~\ref{fig1}).
For a detailed interpretation of our experimental results, we have
performed self-consistent calculations in the multiband
envelope-function approximation \cite{win03} of the valence band
profile of the active region using the nominal growth parameters.
The results of the calculation are plotted in Fig.~\ref{fig2}(b).
The solid (dashed) lines show the HH (LH) subbands and the
strain-split effective potentials for these states. The calculation
predicts a splitting of the lowest HH and LH subbands of about
200~meV. This huge HH-LH splitting is caused by the strong quantum
confinement of the thin Ge layer and the large biaxial strain due to
the lattice mismatch between Ge and Si.

The behavior of the HH states in our device is in sharp contrast to
electron states for which it is well-known that the Zeeman splitting
is proportional to the total magnetic field $B$ irrespective of the
orientation of $B$ relative to the epitaxial layer. Furthermore,
confinement potential and strain do not affect the Zeeman energy of
electron states. In the case of HH states, on the other hand, the
Zeeman splitting in a field $B_\|$ competes with HH-LH splitting;
Zeeman splitting is the smaller the larger the HH-LH splitting and
vice versa \cite{win03}. The appropriate situation can be created in
a narrow QW or by application of uniaxial or biaxial stress. Our
samples satisfy both of these requirements so that we obtain a huge
HH-LH splitting, as can be seen in the band profile in
Fig.~\ref{fig2}(b), resulting in a vanishingly small Zeeman
splitting in a field $B_{\|}$.  This explains why we do not observe
a Zeeman splitting in Fig.~\ref{fig5}.

In the discussion of Fig.~\ref{fig1} only the isotropic part of the
bulk Zeeman term was taken into account \cite{lut56, win03}. This is
the dominant term. The anisotropic part is typically two orders of
magnitude smaller than the isotropic part and the calculations
predict for our structure that it gives rise to a linear splitting
with $\Delta E_\mathrm{HH}^\| \approx 0.18$~meV at $B_\| = 18$~T.
Such a small splitting cannot be resolved in our experiment due to
the width of the conductance peaks. It corresponds to
$g_\mathrm{HH}^\| \approx 0.17$ which is almost two orders of
magnitude smaller than $g_\mathrm{HH}^\perp = 6.3$

In rare cases of conductance peaks we see a slightly different
behavior. In Fig.~\ref{fig5} two conductance maxima at $+10$~mV and
$-10$~mV indicate a small nonlinear splitting above 10~T. If HH-LH
coupling is taken into account, we get a splitting cubic in $B_\|$
and inversely proportional to the HH-LH splitting, $\Delta
E_\mathrm{Z} \propto B_{\|}^{3} / |E_\mathrm{HH} -E_\mathrm{LH}|$
(Ref.~\cite{win03}). For a fully strained system
[Fig.~\ref{fig2}(b)] the calculated splitting due to this term is
even smaller than the splitting due to the anisotropic Zeeman term.
However, it is conceivable that the levels showing a splitting
nonlinear in $B_\|$ are related to shallow acceptors situated in
sample regions of slightly relaxed strain (e.g., close to misfit
dislocations). In these regions the HH-LH splitting is thus reduced
and the cubic Zeeman splitting increases for these levels. This can
explain why a nonlinear splitting is observable for the two
conductance peaks at $\pm 10$~mV, but not for the majority of the
conductance resonances which are due to impurities in highly
strained regions.

In conclusion, we have performed a detailed study of Zeeman
splitting of shallow acceptor levels in a thin Si/Ge/Si quantum
well, by using resonant tunneling spectroscopy. In a magnetic field
orientated perpendicular to the layer a large linear Zeeman
splitting can be observed for magnetic fields up to 18~T. In an
in-plane magnetic field the Zeeman splitting is suppressed. The
giant anisotropy of the Zeeman splitting is a consequence of the
huge heavy hole-light hole splitting produced by a large biaxial
strain and a strong quantum confinement in the narrow Ge quantum
well. It opens a new way to $g$-factor engineering for spintronics
and quantum computing.


\begin{thebibliography}{30}
\bibitem{temprat} The transport measurements in Fig.~\ref{fig5}
    were performed at 1~K which resulted in a full width at half
    maximum of the corresponding conductance peaks of 0.35~meV.
\end{thebibliography}
\end{document}